\documentclass{vldb}
\usepackage{graphicx}
\usepackage{mathtools}
\usepackage{balance}
\usepackage{color}
\usepackage{multirow}
\usepackage{comment}
\usepackage{listings}
\usepackage{enumitem}
\usepackage{amssymb}

\usepackage[hidelinks]{hyperref}

\usepackage{soul}
\setulcolor{red}
\setstcolor{red}

\lstdefinelanguage{pseudo}{
 morekeywords={for,if,then,else,send,to,id,w,tag,val,true,false,inf,
   int,float,compute},
 sensitive=false,
 morecomment=[l]{//},
}

\lstdefinelanguage{palgol}{
 morekeywords={for,in,do,until,fix,repeat,exists,forall,end,if,else,
  let,true,false,fst,snd,to_int,to_float,ref,val,inf,local,remote,V,
  input,output,field,extern,maximum,minimum,sum,and,or,random,
  Nbr,In,Out,Id},
 sensitive=true,
 morecomment=[l]{//},
}
\makeatletter
\newcommand{\shorteq}{%
  \settowidth{\@tempdima}{--}%
  \resizebox{\@tempdima}{\height}{=}%
}
\makeatother
\newcommand{\shorteqq}{\mathop{\shorteq\shorteq}}
\newcommand{\knows}[2]{\mathrm K_{#1}\,{#2}}

\newcommand{\plus}{\raisebox{.25ex}{\scalebox{.8}{+}}}
\newcommand{\plusplus}{\plus\plus}

\begin{document}

\title{Palgol: A High-Level DSL for Vertex-Centric Graph Processing with Remote Data Access}

\numberofauthors{3}

\author{
\alignauthor Yongzhe Zhang\\
 \affaddr{National Institute of Informatics, Tokyo, Japan}\\
 \email{zyz915@nii.ac.jp}
\alignauthor Hsiang-Shang Ko\\
 \affaddr{National Institute of Informatics, Tokyo, Japan}\\
 \email{hsiang-shang@nii.ac.jp}
\alignauthor Zhenjiang Hu\\
 \affaddr{National Institute of Informatics, Tokyo, Japan}\\
 \email{hu@nii.ac.jp}
}

\maketitle

\begin{abstract}

Pregel is a popular distributed computing model for dealing with large-scale graphs.
However, it can be tricky to implement graph algorithms correctly and efficiently in Pregel's vertex-centric model, especially when the algorithm has multiple computation stages, complicated data dependencies, or even communication over dynamic internal data structures.
Some domain-specific languages (DSLs) have been proposed to provide more intuitive ways to implement graph algorithms, but due to the lack of support for \emph{remote access} --- reading or writing attributes of other vertices through references --- they cannot handle the above mentioned dynamic communication, causing a class of Pregel algorithms with fast convergence impossible to implement.

To address this problem, we design and implement Palgol, a more declarative and powerful DSL which supports remote access.
In particular, programmers can use a more declarative syntax called \emph{chain access} to naturally specify dynamic communication as if directly reading data on arbitrary remote vertices.
By analyzing the logic patterns of chain access, we provide a novel algorithm for compiling Palgol programs to efficient Pregel code.
We demonstrate the power of Palgol by using it to implement several practical Pregel algorithms, and the evaluation result shows that the efficiency of Palgol is comparable with that of hand-written code.

\end{abstract}

\section{Introduction}
\label{sec:introduction}

The rapid increase of graph data calls for efficient analysis on massive graphs.
Google's Pregel~\cite{pregel} is one of the most popular frameworks for processing large-scale graphs.
It is based on the bulk-synchronous parallel (BSP) model~\cite{bsp}, and adopts the \emph{vertex-centric} computing paradigm to achieve high parallelism and scalability.
Following the BSP model, a Pregel computation is split into \emph{supersteps} mediated by \emph{message passing.}
Within each superstep, all the vertices execute the same user-defined function \emph{compute()} in parallel, where each vertex can read the messages sent to it in the previous superstep, modify its own state, and send messages to other vertices.
Global barrier synchronization happens at the end of each superstep, delivering messages to their designated receivers before the next superstep.
Despite its simplicity, Pregel has demonstrated its usefulness in implementing many interesting graph algorithms
\cite{pregel,QuWH12,optimizing,XiYZ14,connectivity}.

Despite the power of Pregel, it is a big challenge to implement a graph algorithm correctly and efficiently in it \cite{connectivity}, especially when the algorithm consists of multiple stages and complicated data dependencies.
For such algorithms, programmers need to write an exceedingly complicated \textit{compute()} function as the loop body, which encodes all the stages of the algorithm. 
Message passing makes the code even harder to maintain, because one has to trace where the messages are from and what information they carry in each superstep.
Some attempts have been made to ease Pregel programming by proposing domain-specific languages (DSLs), such as Green-Marl~\cite{green14} and Fregel~\cite{fregel}.
These DSLs allow programmers to write a program in a compositional way to avoid writing a complicated loop body, and provide neighboring data access to avoid explicit message passing.
Furthermore, programs written in these DSLs can be automatically translated to Pregel by fusing the components in the programs into a single loop, and mapping neighboring data access into message passing.
However, for efficient implementation, the existing DSLs impose a severe restriction on data access --- each vertex can only access data on their neighboring vertices.
In other words, they do not support general \emph{remote data access} --- reading or writing attributes of other vertices through references.

Remote data access is, however, important for describing a class of Pregel algorithms that aim to accelerate information propagation (which is a crucial issue in handling graphs with large diameters~\cite{connectivity}) by maintaining a dynamic internal structure for communication.
For instance, a parallel pointer jumping algorithm maintains a tree (or list) structure in a distributed manner by letting each vertex store a reference to its current parent (or predecessor), and during the computation, every vertex constantly exchanges data with the current parent (or predecessor) and modifies the reference to reach the root vertex (or the head of the list).
Such computational patterns can be found in the algorithms like the Shiloach-Vishkin connected component algorithm~\cite{connectivity} (see \autoref{sec:sv-algorithm} for more details), the list ranking algorithm (see \autoref{sec:list-ranking}) and Chung and Condon's minimum spanning forest (MSF) algorithm~\cite{boruvka}.
However, these computational patterns cannot be implemented with only neighboring access, and therefore cannot be expressed in any of the existing high-level DSLs.

It is, in fact, hard to equip DSLs with efficient remote reading.
First, when translated into Pregel's message passing model, remote reads require multiple rounds of communication to exchange information between the reading vertex and the remote vertex, and it is not obvious how the communication cost can be minimized.
Second, remote reads would introduce more involved data dependencies, making it difficult to fuse program components into a single loop.
Things become more complicated when there is \emph{chain access}, where a remote vertex is reached by following a series of references.
Furthermore, it is even harder to equip DSLs with remote writes in addition to remote reads.
For example, Green-Marl detects read/write conflicts, which complicate its programming model; Fregel has a simpler functional model, which, however, cannot support remote writing without major extension.
A more careful design is required to make remote reads and writes efficient and friendly to programmers.

In this paper, we propose a more powerful DSL called Palgol\footnote{Palgol stands for {\bf P}regel {\bf algo}rithmic {\bf l}anguage. The system with all implementation codes and test examples is available at \url{https://bitbucket.org/zyz915/palgol}.} that supports remote data access.
In more detail:
\begin{itemize}
 \item
  We propose a new high-level model for vertex-centric computation, where the concept of \emph{algorithmic supersteps} is introduced as the basic computation unit for constructing vertex-centric computation in such a way that remote reads and writes are ordered in a safe way.

 \item
  Based on the new model, we design and implement Palgol, a more declarative and powerful DSL, which supports both remote reads and writes, and allows programmers to use a more declarative syntax called \emph{chain access} to directly read data on remote vertices.
  For efficient compilation from Palgol to Pregel, we develop a logic system to compile chain access to efficient message passing where the number of supersteps is reduced whenever possible.

 \item
  We demonstrate the power of Palgol by working on a set of representative examples, including the Shiloach-Vishkin connected component algorithm and the list ranking algorithm, which use communication over dynamic data structures to achieve fast convergence.

 \item
  The result of our evaluation is encouraging.
  The efficiency of Palgol is comparable with hand-written code for many representative graph algorithms on practical big graphs, where execution time varies from a $2.53\%$ speedup to a $6.42\%$ slowdown in ordinary cases, while the worst case is less than a $30\%$ slowdown.
\end{itemize}

The rest of the paper is organized as follows.
\autoref{sec:core} introduces algorithmic supersteps and the essential parts of Palgol, \autoref{sec:compilation} presents the compiling algorithm, and \autoref{sec:evaluation} presents evaluation results.
Related work is discussed in \autoref{sec:related}, and \autoref{sec:conclusions} concludes this paper with some outlook.

\section{The Palgol Language}
\label{sec:core}

This section first introduces a high-level vertex-centric programming model~(\autoref{sec:model}), in which an algorithm is decomposed into atomic vertex-centric computations and high-level combinators, and a vertex can access the entire graph through the references it stores locally.
Next we define the Palgol language based on this model, and explain its syntax and semantics~(\autoref{sec:syntax}).
Then, we show how to write the classic shortest path algorithm in Palgol.
Finally we use two representative examples --- the Shiloach-Vishkin connected component algorithm~(\autoref{sec:sv-algorithm}) and the list ranking algorithm~(\autoref{sec:list-ranking}) --- to demonstrate how Palgol can concisely describe vertex-centric algorithms with dynamic internal structures using remote access.

\subsection{The High-Level Model}
\label{sec:model}

The high-level model we propose uses remote reads and writes instead of message passing to allow programmers to describe vertex-centric computation more intuitively.
Moreover, the model remains close to the Pregel computation model, in particular keeping the vertex-centric paradigm and barrier synchronization, making it possible to automatically derive a valid and efficient Pregel implementation from an algorithm description in this model, and in particular arrange remote reads and writes without data conflicts.

In our high-level model, the computation is constructed from some basic components which we call \emph{algorithmic supersteps}.
An algorithmic superstep is a piece of vertex-centric computation which takes a graph containing a set of vertices with local states as input, and outputs the same set of vertices with new states.
Using algorithmic supersteps as basic building blocks, two high-level operations \emph{sequence} and \emph{iteration} can be used to glue them together to describe more complex vertex-centric algorithms that are iterative and/or consist of multiple computation stages:
the \emph{sequence} operation concatenates two algorithmic supersteps by taking the result of the first step as the input of the second one, and the \emph{iteration} operation repeats a piece of vertex-centric computation until some termination condition is satisfied.

\begin{figure}[t]
 \centering
 \includegraphics[width=0.45\textwidth]{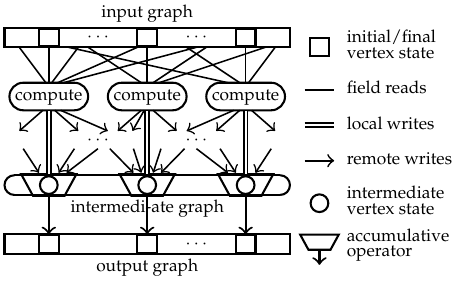}
 \caption{In an algorithmic superstep, every vertex performs local computation (including field reads and local writes) and remote updating in order.}
 \label{fig:algostep}
 \vspace{-2.5ex}
\end{figure}

The distinguishing feature of algorithmic supersteps is remote access.
Within each algorithmic superstep (illustrated in \autoref{fig:algostep}), all vertices compute in parallel, performing the same computation specified by programmers.
A vertex can read the fields of any vertex in the input graph; it can also write to arbitrary vertices to modify their fields, but the writes are performed on a separate graph rather than the input graph (so there are no read-write conflicts).
We further distinguish \emph{local writes} and \emph{remote writes} in our model:
local writes can only modify the current vertex's state, and are first performed on an intermediate graph (which is initially a copy of the input graph);
next, remote writes are propagated to the destination vertices to further modify their intermediate states.
Here, a remote write consists of a remote field, a value and an ``accumulative'' assignment (like \texttt{+=} and \texttt{|=}), and that field of the destination vertex is modified by executing the assignment with the value on its right-hand side.
We choose to support only accumulative assignments so that the order of performing remote writes does not matter.

More precisely, an algorithmic superstep is divided into the following two phases:
\begin{itemize}
 \item a \emph{local computation} (LC) phase, in which a copy of the input graph is created as the intermediate graph, and then each vertex can read the state of any vertex in the input graph, perform local computation, and modify its own state in the intermediate graph, and
 \item a \emph{remote updating} (RU) phase, in which each vertex can modify the states of any vertices in the intermediate graph by sending remote writes.
After processing all remote writes are processed, the intermediate graph is returned as the output graph.
\end{itemize}
Among these two phases, the RU phase is optional, in which case the intermediate graph produced by the LC phase is used directly as the final result.

\subsection{An Overview of Palgol}
\label{sec:syntax}

We present our DSL Palgol next, whose design follows the high-level model we introduced in the previous subsection.
Figure~\ref{fig:syntax-simplified} shows the essential part of the syntax of Palgol.
As described by the syntactic category \textit{step}, an algorithmic superstep in Palgol is a code block
enclosed by ``\textbf{for} \textit{var} \textbf{in} \textbf{V}'' and ``\textbf{end}'', where \textit{var} is a variable name that can be used in the code block for referring to the current vertex (and \textbf{V} stands for the set of vertices of the input graph).
Such steps can then be composed (by sequencing) or iterated until a termination condition is met (by enclosing them in ``\textbf{do}'' and ``\textbf{until} \ldots'').
Palgol supports several kinds of termination condition, but in this paper we focus on only one kind of termination condition called \textit{fixed point}, since it is extensively used in many algorithms.
The semantics of fixed-point iteration is iteratively running the program enclosed by \textbf{do} and \textbf{until}, until the specified fields stabilize.

\begin{figure}[t]
\normalsize
\[
\begin{array}{lcl}
\mathit{int} & = & \hbox{integer} \\
\mathit{float} & = & \hbox{floating-point number} \\
\mathit{var} & = & \hbox{identifier starting with lowercase letter} \\
\mathit{field} & = & \hbox{identifier starting with capital letter} \\
\\
\mathit{prog}  & \Coloneqq & \mathit{step}~|~\mathit{prog_1}\ldots\mathit{prog_n}~|~\mathit{iter} \\
\mathit{iter} & \Coloneqq & \mathbf{do}~\langle~\mathit{prog}~\rangle~\mathbf{until}~\mathbf{fix}~[~\mathit{field_1},\ldots,\mathit{field_n}~] \\
\mathit{step}& \Coloneqq & \mathbf{for}~\mathit{var}~\mathbf{in}~\mathbf{V}~\langle~\mathit{block}~\rangle~\mathbf{end} \\
\mathit{block} & \Coloneqq & \mathit{stmt_1} \ldots \mathit{stmt_n} \\
\mathit{stmt}  & \Coloneqq & \mathbf{if}~\mathit{exp}~\langle~\mathit{block}~\rangle~|~\mathbf{if}~\mathit{exp}~\langle~\mathit{block}~\rangle~\mathbf{else}~\langle~\mathit{block}~\rangle \\
 & | & \mathbf{for}~(\mathit{var}\leftarrow\mathit{exp})~\langle~\mathit{block}~\rangle \\
 & | & \mathbf{let}~\mathit{var}=\mathit{exp} \\
 & | & \mathbf{local}\mathit{_{opt}}~\mathit{field}~[~\mathit{var}~]~\mathit{op_{local}}~\mathit{exp} \\
 & | & \mathbf{remote}~\mathit{field}~[~\mathit{exp}~]~\mathit{op_{remote}}~\mathit{exp} \\
\mathit{exp}   & \Coloneqq & \mathit{int}~|~\mathit{float}~|~\mathit{var}~|~\mathbf{true}~|~\mathbf{false}~|~\mathbf{inf} \\
 & | & \mathbf{fst}~\mathit{exp}~|~\mathbf{snd}~\mathit{exp}~|~(\mathit{exp},\mathit{exp}) \\
 & | & \mathit{exp}.\mathbf{ref}~|~\mathit{exp}.\mathbf{val}~|~\{\mathit{exp},\mathit{exp}\}~|~\{\mathit{exp}\} \\
 & | & \mathit{exp}~?~\mathit{exp}:\mathit{exp}~|~(~\mathit{exp}~)~|~\mathit{exp}~\mathit{op_{b}}~\mathit{exp}~|~\mathit{op_{u}}~\mathit{exp} \\
 & | & \mathit{field}~[~\mathit{exp}~] \\
 & | & \mathit{func_{opt}}~[~\mathit{exp}~|~\mathit{var}\leftarrow\mathit{exp},\mathit{exp_1},\ldots,\mathit{exp_n}~] \\
\mathit{func}  & \Coloneqq & \mathbf{maximum}~|~\mathbf{minimum}~|~\mathbf{sum}~|~\ldots \\
 \end{array}
\]
\caption{Essential part of Palgol's syntax. Palgol is indentation-based, and two special tokens `$\langle$' and `$\rangle$' are introduced to delimit indented blocks.}
\label{fig:syntax-simplified}
\end{figure}

Corresponding to an algorithmic superstep's remote access capabilities, in Palgol we can read a field of an arbitrary vertex using a global field access expression of the form $\mathit{field}~[\,\mathit{exp}\,]$, 
where $\mathit{field}$ is a user-specified field name and $\mathit{exp}$ should evaluate to a vertex id.
Such expression can be updated by local or remote assignments, where an assignment to a remote vertex should always be accumulative and prefixed with the keyword $\mathbf{remote}$.
One more thing about remote assignments is that they take effect only in the RU phase (after the LC phase), regardless of where they occur in the program.

There are some predefined fields that have special meaning in our language.
$\mathbf{Nbr}$ is the edge list in undirected graphs, and $\mathbf{In}$ and $\mathbf{Out}$ respectively store incoming and outgoing edges for directed graphs.
Essentially, these are normal fields of a predefined type for representing edges, and most importantly, the compiler assumes a form of symmetry on these fields (namely that every edge is stored consistently on both of its end vertices), and uses the symmetry to produce more efficient code.
Then, $\mathbf{Id}$ is an immutable field that stores the vertex identifier for each vertex (required by the Pregel framework), whose type is user-specified but currently we can simply treat it as an integer.

The rest of the syntax for Palgol steps is similar to an ordinary programming language.
Particularly, we introduce a specialized pair type (expressions in the form of $\{\mathit{exp},\mathit{exp}\}$) for representing a reference with its corresponding value (e.g., an edge in a graph), and use $.\mathbf{ref}$ and $.\mathbf{val}$ respectively to access the reference and the value respectively, to make the code easy to read.
Some functional programming constructs are also used here, like let-binding and list comprehension.
There is also a foreign function interface that allows programmers to invoke functions written in a general-purpose language, but we omit the detail from the paper.

\subsection{Single-Source Shortest Path Algorithm}
\label{sec:sssp}

The single-source shortest path problem is among the best known in graph theory and arises in a wide variety of applications.
The idea of this algorithm is fairly simple, which is an iterative computation until the following equation holds:
$$ \mathit{dist}[v] =\begin{cases}0 & \text{$v$~is the source} \\ \min_{u\in \mathit{In}(v)}~(\mathit{dist}[u]~+~\mathit{len}(v,~u)) & \text{otherwise} \end{cases} $$
We can concisely capture the essence of the shortest path algorithm in a Palgol program, as shown in Figure~\ref{fig:sssp-palgol}.
In this program, we store the distance of each vertex from the source in the $D$~field, and use a boolean field~$A$ to indicate whether the vertex is active.
There are two steps in this program.
In the first step (lines 1--4), every vertex initializes its own distance and the $A$~field.
Then comes the iterative step (lines 6--13) inside $\mathbf{do} \ldots \mathbf{until}~\mathbf{fix}~[D]$, which runs until every vertex's distance stabilizes.
Using a list comprehension (lines 7--8), each vertex iterates over all its active incoming neighbors (those whose $A$~field is true), and generates a list containing the sums of their current distances and the corresponding edge weights.
More specifically, the list comprehension goes through every edge~$e$ in the incoming edge list $\mathbf{In}\,[u]$ such that $A\,[e.\mathbf{ref}]$ is true, and puts $D\,[e.\mathbf{ref}] + e.\mathbf{val}$ in the generated list, where $e.\mathbf{ref}$ represents the neighbor's vertex id and $e.\mathbf{val}$ the edge weight.
Finally, we pick the minimum value from the generated list as \textit{minD}, and update the local fields.

\begin{figure}[thp]
\begin{lstlisting}[basicstyle=\footnotesize\ttfamily]
for u in V
  D[u] := (Id[u] == 0 ? 0 : inf)
  A[u] := (Id[u] == 0)
end
do
  for u in V
    let minD = minimum[ D[e.ref] + e.val
             | e <- In[u], A[e.ref] ]
    A[u] := false
    if (minD < D[u])
      A[u] := true
      D[u] := minD
  end
until fix[D]
\end{lstlisting}
\vspace{-2ex}
\caption{The SSSP program in Palgol}
\label{fig:sssp-palgol}
\end{figure}

\subsection{The Shiloach-Vishkin Connected Component Algorithm}
\label{sec:sv-algorithm}

Here is our first representative Palgol example: the \emph{Shiloach-Vishkin (S-V) connected component algorithm}~\cite{connectivity}, which can be expressed as the Palgol program in \autoref{fig:svppa-code}.
A traditional HashMin connected component algorithm~\cite{connectivity} based on neighborhood access takes time proportional to the input graph's diameter, which can be large in real-world graphs.
In contrast, the S-V algorithm can calculate the connected components of an undirected graph in a logarithmic number of supersteps; to achieve this fast convergence, the capability of accessing data on non-neighboring vertices is essential.

In the S-V algorithm, the connectivity information is maintained using the classic disjoint set data structure~\cite{disjointset}.
Specifically, the data structure is a forest, and vertices in the same tree are regarded as belonging to the same connected component.
Each vertex maintains a parent pointer that either points to some other vertex in the same connected component, or points to itself, in which case the vertex is the root of a tree.
We henceforth use $D[u]$ to represent this pointer for each vertex $u$.
The S-V algorithm is an iterative algorithm that begins with a forest of $n$ root nodes, and in each step it tries to discover edges connecting different trees and merge the trees together.
In a vertex-centric way, every vertex~$u$ performs one of the following operations depending on whether its parent $D[u]$ is a root vertex:
\begin{itemize}
 \item \textbf{tree merging:}
  if $D[u]$ is a root vertex, then $u$ chooses one of its neighbors' current parent (to which we give a name $t$), and makes $D[u]$ point to $t$ if $t<D[u]$ (to guarantee the correctness of the algorithm).
  When having multiple choices in choosing the neighbors' parent $p$, or when different vertices try to modify the same parent vertex's pointer, the algorithm always uses the ``minimum'' as the tiebreaker for fast convergence.
 \item \textbf{pointer jumping:}
  if $D[u]$ is not a root vertex, then $u$ modifies its own pointer to its current ``grandfather'' ($D[u]$'s current pointer).
  This operation reduces $u$'s distance to the root vertex, and will eventually make~$u$ a direct child of the root vertex so that it can perform the above tree merging operation.
\end{itemize}
The algorithm terminates when all vertices' pointers do not change after an iteration, in which case all vertices point to some root vertex and no more tree merging can be performed.
Readers interested in the correctness of this algorithm are referred to the original paper~\cite{connectivity} for more details.

The implementation of this algorithm is complicated, which contains roughly 120 lines of code\footnote{\url{http://www.cse.cuhk.edu.hk/pregelplus/code/apps/basic/svplus.zip}} for the $\textit{compute()}$ function alone.
Even for detecting whether the parent vertex $D[u]$ is a root vertex for each vertex $u$, it has to be translated into three supersteps containing a query-reply conversation between each vertex and its parent.
In contrast, the Palgol program in Figure \ref{fig:svppa-code} can describe this algorithm concisely in 13 lines, due to the declarative remote access syntax.
This piece of code contains two steps, where the first one (lines 1--3) performs simple initialization, and the other (lines 5--12) is inside an iteration as the main computation.
We also use the field~$D$ to store the pointer to the parent vertex.
Let us focus on line~6, which checks whether $u$'s parent is a root.
Here we simply check $D[D[u]] \shorteqq D[u]$, i.e., whether the pointer of the parent vertex $D[D[u]]$ is equal to the parent's id $D[u]$.
This expression is completely declarative, in the sense that we only specify what data is needed and what computation we want to perform, instead of explicitly implementing the message passing scheme.

\begin{figure}[thp]
\begin{lstlisting}[basicstyle=\footnotesize\ttfamily]
for u in V
  local D[u] := u
end
do
  for u in V
    if (D[D[u]] == D[u])
      let t = minimum [ D[e.id]
            | e <- Nbr[u] ]
      if (t < D[u])
        remote D[D[u]] <?= t
    else
      local D[u] := D[D[u]]
  end
until fix[D]
\end{lstlisting}
\vspace{-2ex}
\caption{The S-V program}
\label{fig:svppa-code}
\end{figure}

The rest of the algorithm can be straightforwardly associated with the Palgol program.
If $u$'s parent is a root, we generate a list containing all neighboring vertices' parent id ($D[e.\mathbf{ref}]$), and then bind the minimum one to the variable~$t$ (line~7).
Now $t$~is either \textbf{inf} if the neighbor list is empty or a vertex id; in both cases we can use it to update the parent's pointer (lines 9--10) via a remote assignment.
One important thing is that the parent vertex ($D[u]$) may receive many remote writes from its children, where only one of the children providing the minimum $t$ can successfully perform the updating.
Here, the statement \texttt{a <?= b} is an accumulative assignment, whose meaning is the same as \texttt{a := min(a, b)}.
Finally, for the $\mathbf{else}$ branch, we (locally) assign $u$'s grandparent's id to $u$'s $D$ field.

\subsection{The List Ranking Algorithm}
\label{sec:list-ranking}

Another example is the \emph{list ranking} algorithm, which also needs communication over a dynamic structure during computation.
Consider a linked list~$L$ with $n$~elements, where each element $u$ stores a value $\mathit{val}(u)$ and a link to its predecessor $\mathit{pred}(u)$.
At the head of $L$ is a virtual element~$v$ such that $\mathit{pred}(v)=v$ and $\mathit{val}(v)=0$.
For each element~$u$ in~$L$, define $\mathit{sum}(u)$ to be the sum of the values of all the elements from~$u$ to the head (following the predecessor links).
The list ranking problem is to compute $\mathit{sum}(u)$ for each element $u$.
If $\mathit{val}(u)=1$ for every vertex~$u$ in~$L$, then $\mathit{sum}(u)$ is simply the rank of~$u$ in the list.
List ranking can be solved using a typical pointer-jumping algorithm in parallel computing with a strong performance guarantee.
Yan~et~al.~\cite{connectivity} demonstrated how to compute the pre-ordering numbers for all vertices in a tree in $O(\log n)$ supersteps using this algorithm, as an internal step to compute bi-connected components (BCC).%
\footnote{BCC is a complicated algorithm, whose efficient implementation requires constructing an intermediate graph, which is currently beyond Palgol's capabilities. Palgol is powerful enough to express the rest of the algorithm, however.}

\begin{figure}[t]
\begin{lstlisting}[basicstyle=\footnotesize\ttfamily]
for u in V
  Sum[u] := Val[u]
end
do
  for u in V
    if (Pred[Pred[u]] != Pred[u])
      Sum[u] += Sum[Pred[u]]
      Pred[u] := Pred[Pred[u]]
  end 
until fix[Pred]
\end{lstlisting}
\vspace{-2ex}
\caption{The list ranking program}
\label{fig:ranking-code}
\end{figure}

We give the Palgol implementation of list ranking in~\autoref{fig:ranking-code} (which is a 10-line program, whereas the Pregel implementation\footnote{\url{http://www.cse.cuhk.edu.hk/pregelplus/code/apps/basic/bcc.zip}} contains around 60~lines of code).
$\mathit{Sum}[u]$ is initially set to\,$\mathit{Val}[u]$ for every~$u$ at line~2; inside the fixed-point iteration (lines 5--9), every~$u$ moves $\mathit{Pred}[u]$ toward the head of the list and updates $\mathit{Sum}[u]$ to maintain the invariant that $\mathit{Sum}[u]$ stores the sum of a sublist from itself to the successor of $\mathit{Pred}[u]$.
Line~6 checks whether $u$~points to the virtual head of the list, which is achieved by checking $\mathit{Pred}[\mathit{Pred}[u]] \shorteqq \mathit{Pred}[u]$, i.e., whether the current predecessor $\mathit{Pred}[u]$ points to itself.
If the current predecessor is not the head, we add the sum of the sublist maintained in $\mathit{Pred}[u]$ to the current vertex~$u$, by reading $\mathit{Pred}[u]$'s $\mathit{Sum}$ and $\mathit{Pred}$ fields and modifying $u$'s own fields accordingly.
Note that since all the reads are performed on a snapshot of the input graph and the assignments are performed on an intermediate graph, there is no need to worry about data dependencies.

\subsection{Vertex Inactivation}

In some Pregel algorithms, we may want to inactivate vertices during computation.
Typical examples include some matching algorithms like randomized bipartite matching~\cite{pregel} and approximate maximum weight matching~\cite{optimizing}, where matched vertices are no longer needed in subsequent computation, and the minimum spanning forest algorithm~\cite{optimizing} where the graph gradually shrinks during computation.

In Palgol, we model the behavior of inactivating vertices as a special Palgol step, which can be freely composed with other Palgol programs.
The syntactic category of \textit{step} is now defined as follows:
\[
\begin{array}{lcl}
\mathit{step} & \Coloneqq & \mathbf{for}~\mathit{var}~\mathbf{in}~\mathbf{V}~\langle~\mathit{block}~\rangle~\mathbf{end} \\
 & | & \mathbf{stop}~\mathit{var}~\mathbf{where}~\mathit{exp} \\
\end{array}
\]
The special Palgol step stops those vertices satisfying the condition specified by the boolean-valued expression \textit{exp}, which can refer to the current vertex $\mathit{var}$.
The semantics of stopping vertices is different from Pregel's voting to halt mechanism.
In Pregel, an inactive vertex can be activated by receiving messages, but such semantics is unsuitable for Palgol, since we already hide message passing from programmers.
Instead, a stopped vertex in Palgol will become immutable and never perform any subsequent local computation, but other vertices can still access its fields.
This feature is still experimental and we do not further discuss it in this paper; it is, however, essential for achieving the performance reported in \autoref{sec:evaluation}.

\section{Compiling Palgol to Pregel}
\label{sec:compilation}

In this section, we present the compiling algorithm to transform Palgol to Pregel.
The task overall is complicated and highly technical, but the main problems are the following two: how to translate Palgol steps into Pregel supersteps, and how to implement sequence and iteration, which will be presented in \autoref{sec:trans-step} and \autoref{sec:trans-iter} respectively.
When compiling a single Palgol step, the most challenging part is the remote reads, for which we first give a detailed explanation in \autoref{sec:trans-read}.
We also mention an optimization based on Pregel's combiners in \autoref{sec:combiner}.

\subsection{Compiling Remote Reads}
\label{sec:trans-read}

In current Palgol, our compiler recognizes two forms of remote reads.
The first one is called \emph{consecutive field access} (or \emph{chain access} for short), which uses nested field access expressions to acquire remote data.
The second one is called \emph{neighborhood access} where a vertex may use chain access to acquire data from \emph{all} its neighbors, and this can be described using the list comprehension (e.g., line~7 in \autoref{fig:svppa-code}) or for-loop syntax in Palgol.
The combination of these two remote read patterns is already sufficient to express quite a wide range of practical Pregel algorithms according to our experience.
In this section, we present the key algorithms to compile these two remote read patterns to message passing in Pregel.

\subsubsection{Consecutive Field Access Expressions}
\label{sec:consecutive}

\textbf{Definition and challenge of compiling}:
Let us begin from the first case of remote reads, which is consecutive field access expressions (or chain access) starting from the current vertex.
As an example, supposing that the current vertex is~$u$, and $D$~is a field for storing a vertex id, then $D[D[u]]$ is a consecutive field access expression, and so is $D[D[D[D[u]]]]$ (which we abbreviate to $D^4[u]$ in the rest of this section).
Generally speaking, there is no limitation on the depth of a chain access or the number of fields involved in the chain access.

As a simple example of the compilation, to evaluate $D[D[u]]$ on every vertex~$u$, a straightforward scheme is a request-reply conversation which takes two rounds of communication:
in the first superstep, every vertex~$u$ sends a request to (the vertex whose id is) $D[u]$ and the request message should contain $u$'s own id;
then in the second superstep, those vertices receiving the requests should extract the sender's ids from the messages, and reply its $D$ field to them.

When the depth of such chain access increases, it is no longer trivial to find an efficient scheme, where efficiency is measured in terms of the number of supersteps taken.
For example, to evaluate $D^4[u]$ on every vertex $u$, a simple query-reply method takes six rounds of communication by evaluating $D^2[u]$, $D^3[u]$ and $D^4[u]$ in turn, each taking two rounds, but the evaluation can actually be done in only three rounds with our compilation algorithm, which is not based on request-reply conversations.

\textbf{Logic system for compiling chain access}:
The key insight leading to our compilation algorithm is that we should consider not only the expression to evaluate but also the vertex on which the expression is evaluated.
To use a slightly more formal notation (inspired by Halpern and Moses~\cite{Halpern-common-knowledge}), we write $\forall u.\,\knows{v(u)}{e(u)}$, where $v(u)$~and~$e(u)$ are chain access expressions starting from~$u$, to describe the state where every vertex $v(u)$ ``knows'' the value of the expression $e(u)$; then the goal of the evaluation of $D^4[u]$ can be described as $\forall u.\,\knows{u}{D^4[u]}$.
Having introduced the notation, the problem can now be treated from a logical perspective, where we aim to search for a derivation of a target proposition from a few axioms.

There are three axioms in our logic system:
\begin{enumerate}
\item $\forall u.\,\knows{u}{u}$
\item $\forall u.\,\knows{u}{\mathit{D}[u]}$
\item $(\forall u.\,\knows{w(u)}{e(u)}) \wedge (\forall u.\,\knows{w(u)}{v(u)}) \implies \forall u.\,\knows{v(u)}{e(u)}$
\end{enumerate}
The first axiom says that every vertex knows its own id, and the second axiom says every vertex can directly access its local field $D$.
The third axiom encodes message passing: if we want every vertex $v(u)$ to know the value of the expression $e(u)$, then it suffices to find an intermediate vertex $w(u)$ which knows both the value of $e(u)$ and the id of $v(u)$, and thus can send the value to $v(u)$.
As an example, Figure~\ref{fig:d4u-rules} shows the solution generated by our algorithm to solve $\forall u.\,\knows{u}{D^4[u]}$, where each line is an instance of the message passing axiom.

\begin{figure}[t]
\vspace{-1.5ex}
\normalsize
\begin{align*}
(\forall u.\,\knows{\mathrlap{u}\phantom{D^2[u]}}{\mathrlap{u}\phantom{D^2[u]}}) \mathrel\wedge (\forall u.\,\knows{\mathrlap{u}\phantom{D^2[u]}}{\mathrlap{D[u]}\phantom{D^2[u]}}) &\implies \forall u.\,\knows{\mathrlap{D[u]}\phantom{D^2[u]}}{u} \\
(\forall u.\,\knows{\mathrlap{D[u]}\phantom{D^2[u]}}{\mathrlap{u}\phantom{D^2[u]}}) \mathrel\wedge (\forall u.\,\knows{\mathrlap{D[u]}\phantom{D^2[u]}}{D^2[u]}) &\implies \forall u.\,\knows{\mathrlap{D^2[u]}\phantom{D^2[u]}}{u} \\
(\forall u.\,\knows{\mathrlap{D[u]}\phantom{D^2[u]}}{D^2[u]}) \mathrel\wedge (\forall u.\,\knows{\mathrlap{D[u]}\phantom{D^2[u]}}{\mathrlap{u}\phantom{D^2[u]}}) &\implies \forall u.\,\knows{\mathrlap{u}\phantom{D^2[u]}}{D^2[u]} \\
(\forall u.\,\knows{\mathrlap{D^2[u]}\phantom{D^2[u]}}{D^4[u]}) \mathrel\wedge (\forall u.\,\knows{\mathrlap{D^2[u]}\phantom{D^2[u]}}{\mathrlap{u}\phantom{D^2[u]}}) &\implies \forall u.\,\knows{\mathrlap{u}\phantom{D^2[u]}}{D^4[u]}
\end{align*}
\caption{A derivation of $\forall u.\,\knows{u}{D^4[u]}$}
\label{fig:d4u-rules}
\vspace{-2ex}\end{figure}

\begin{figure}[t]
 \centering
 \includegraphics[width=0.45\textwidth]{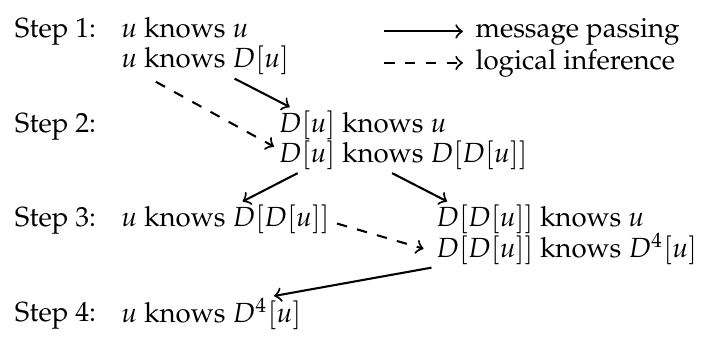}
 \caption{Interpretation of the derivation of $\forall u.\,\knows{u}{D^4[u]}$}
 \label{fig:d4u-msg}
\vspace{-2ex}\end{figure}

\autoref{fig:d4u-msg} is a direct interpretation of the implications in \autoref{fig:d4u-rules}.
To reach $\forall u.\,\knows{u}{D^4[u]}$, only three rounds of communication are needed.
Each solid arrow represents an invocation of the message passing axiom in Figure~\ref{fig:d4u-rules}, and the dashed arrows represent two logical inferences, one from $\forall u.\,\knows{u}{D[u]}$ to $\forall u.\,\knows{D[u]}{D^2[u]}$ and the other from $\forall u.\,\knows{u}{D^2[u]}$ to $\forall u.\,\knows{D^2[u]}{D^4[u]}$.

The derivation of $\forall u.\,\knows{u}{D^4[u]}$ is not unique, and there are derivations that correspond to inefficient solutions --- for example, there is also a derivation for the six-round solution based on request-reply conversations.
However, when searching for derivations, our algorithm will minimize the number of rounds of communication, as explained below.

\textbf{The compiling algorithm}:
The algorithm starts from a proposition $\forall u.\,\knows{v(u)}{e(u)}$.
The key problem here is to choose a proper $w(u)$ so that, by applying the message passing axiom backwards, we can get two potentially simpler new target propositions $\forall u.\,\knows{w(u)}{e(u)}$ and $\forall u.\,\knows{w(u)}{v(u)}$ and solve them respectively.
The range of such choices is in general unbounded, but our algorithm considers only those simpler than $v(u)$ or $e(u)$.
More formally, we say that $a$~is a \emph{subpattern} of~$b$, written $a \preceq b$, exactly when $b$~is a consecutive field access expression starting from~$a$.
For example, $u$~and $D[u]$ are subpatterns of $D[D[u]]$, while they are all subpatterns of $D^3[u]$.
The range of intermediate vertices we consider is then $\mathrm{Sub}(e(u), v(u))$, where $\mathrm{Sub}$ is defined by
\[ \mathrm{Sub}(a, b) = \{\,c \mid c \preceq a \mathrel\text{or} c \prec b \,\} \]
We can further simplify the new target propositions with the following function before solving them:
\[
\mathit{generalize}(\forall u.\,\knows{a(u)}{b(u)}) =
\begin{cases}
\forall u.\,\knows{u}{(b(u)/a(u))} & \text{if~} a(u) \preceq b(u) \\
\forall u.\,\knows{a(u)}{b(u)} & \text{otherwise}
\end{cases}
\]
where $b(u)/a(u)$ denotes the result of replacing the innermost $a(u)$ in $b(u)$ with~$u$.
(For example, $A[B[C[u]]]/C[u] = A[B[u]]$.)
This is justified because the original proposition can be instantiated from the new proposition.
(For example, $\forall u.\,\knows{C[u]}{A[B[C[u]]]}$ can be instantiated from $\forall u.\,\knows{u}{A[B[u]]}$.)

It is now possible to find an optimal solution with respect to the following inductively defined function $\mathit{step}$, which calculates the number of rounds of communication for a proposition:
\[ \setlength{\arraycolsep}{.2em}\begin{array}{lcl}
\mathit{step}(\forall u.\,\knows{u}{u}) &=& 0 \\
\mathit{step}(\forall u.\,\knows{u}{\mathit{D}[u]}) &=& 0 \\
\mathit{step}(\forall u.\,\knows{v(u)}{e(u)}) &=& \displaystyle 1+ \min_{w(u)\in \mathrm{Sub}(e(u),v(u))}~\max(x, y) \\
\multicolumn{3}{l}{\quad\text{where}~x = \mathit{step}(\mathit{generalize}(\forall u.\,\knows{w(u)}{e(u)}))} \\
\multicolumn{3}{l}{\quad\phantom{\text{where}}~y = \mathit{step}(\mathit{generalize}(\forall u.\,\knows{w(u)}{v(u)}))}
\end{array} \]
It is straightforward to see that this is an optimization problem with optimal and overlapping substructure, which we can solve efficiently with memoization techniques.

With this powerful compiling algorithm, we are now able to handle any chain access expressions.
Furthermore, this algorithm optimizes the generated Pregel program in two aspects.
First, this algorithm derives a message passing scheme with a minimum number of supersteps, thus reduces unnecessary cost for launching supersteps in Pregel framework.
Second, by extending the memoization technique, we can ensure that a chain access expression will be evaluated exactly once even if it appears multiple times in a Palgol step, avoiding redundant message passing for the same value.

\subsubsection{Neighborhood Access}
\label{sec:neighboring-access}

Neighborhood access is another important communication pattern widely used in Pregel algorithms.
Precisely speaking, neighborhood access refers to those chain access expressions inside a non-nested loop traversing an edge list (\textbf{Nbr}, \textbf{In} or \textbf{Out}), where the chain access expressions start from the neighboring vertex.
The following code is a typical example of neighborhood access, which is a list comprehension used in the S-V algorithm program (Figure~\ref{fig:svppa-code}):
\begin{lstlisting}[basicstyle=\footnotesize\ttfamily,firstnumber=7]
    let t = minimum [ D[e.ref]
          | e <- Nbr[u] ]
\end{lstlisting}
Syntactically, a field access expression $D[e.\mathbf{ref}]$ can be easily identified as a neighborhood access.

The compilation of such data access pattern is based on the symmetry that if \emph{all} vertices need to fetch the same field of their neighbors, that will be equivalent to making all vertices send the field to all their neighbors.
This is a well-known technique that is also adopted by Green-Marl and Fregel, so we do not go into the details and simply summarize the compilation procedure as follows:
\begin{enumerate}
 \item
  In the first superstep, we prepare the data from neighbors' perspective.
  Field access expressions like $D[e.\mathbf{ref}]$ now become neighboring vertices' local fields $D[u]$.
  Every vertex then sends messages containing those values to all its neighboring vertices.
 \item
  In the next step, every vertex scans the message list to obtain all the values of neighborhood access, and then executes the loop according to the Palgol program.
\end{enumerate}

\subsection{Compiling Palgol Steps}
\label{sec:trans-step}

\begin{figure}
 \centering
 \includegraphics[width=0.45\textwidth]{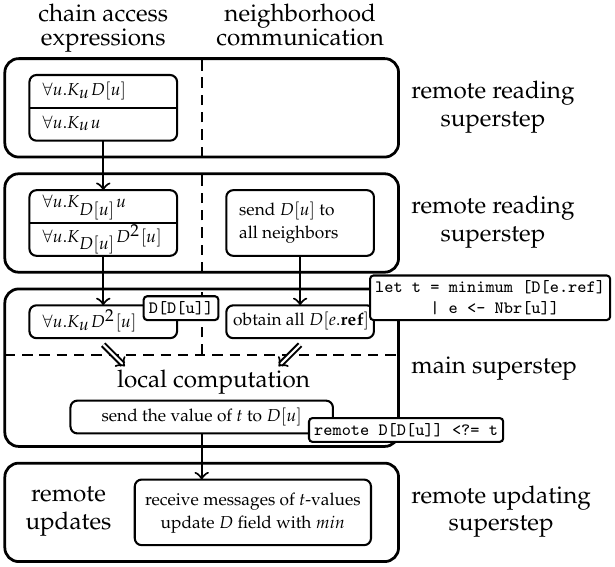}
 \caption{Compiling a Palgol step to Pregel supersteps.}
 \label{fig:compiling-Palgol-step}
\vspace{-2ex}\end{figure}

Having introduced the compiling algorithm for remote data reads in Palgol, here we give a general picture of the compilation for a single Palgol step, as shown in \autoref{fig:compiling-Palgol-step}.
The computational content of every Palgol step is compiled into a \emph{main superstep}.
Depending on whether there are remote reads and writes, there may be a number of \emph{remote reading supersteps} before the main superstep, and a \emph{remote updating superstep} after the main superstep.

We will use the main computation step of the S-V program (lines 5--13 in \autoref{fig:svppa-code}) as an illustrative example for explaining the compilation algorithm, which consists of the following four steps:
\begin{enumerate}

\item We first handle neighborhood access, which requires a sending superstep that provides all the remote data for the loops from the neighbors' perspective. (for S-V algorithm, sending their $D$~field to all their neighbors).
This sending superstep is inserted as a remote reading superstep immediately before the main superstep.

\item We analyze the chain access expressions appearing in the Palgol step with the algorithm in \autoref{sec:trans-read}, and corresponding remote reading supersteps are inserted in the front.
(For the S-V algorithm, the only interesting chain access expression is $D[D[u]]$, which induces two remote reading supersteps realizing a request-reply conversation.)

\item Having handled all remote reads, the main superstep receives all the values needed and proceeds with the local computation.
Since the local computational content of a Palgol step is similar to an ordinary programming language, the transformation is straightforward.

\item What remain to be handled are the remote assignments, which require sending the updating values as messages to the target vertices in the main superstep.
(For S-V algorithm, there is one remote updating statement at line 10, requiring that the value of~$t$ be sent to $D[u]$.)
Then an additional remote updating superstep is added after the main superstep; this additional superstep reads these messages and updates each field using the corresponding remote updating operator.
\end{enumerate}

\subsection{Compiling Sequences and Iterations}
\label{sec:trans-iter}

We finally tackle the problem of compiling sequence and iteration, to assemble Palgol steps into larger programs.

A Pregel program generated from Palgol code is essentially a \emph{state transition machine} (STM) combined with computation code for each state.
In the simplest case, every Palgol step is translated into a ``linear'' STM consisting of a chain of states corresponding to the supersteps like those shown in \autoref{fig:compiling-Palgol-step}.
In general, a generated STM may be depicted as:
\begin{figure}[h]
 \centering
 \includegraphics[width=0.15\textwidth]{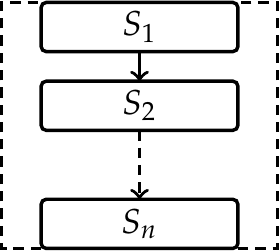}
\end{figure}\\
where there are a start state and an end state, between which there can be more states and transitions, not necessarily having the linear structure.

\subsubsection{Compiling Sequences}

A sequence of two Palgol programs uses the first program to transform an initial graph to an intermediate one, which is then transformed to a final graph using the second program.
To compile the sequence, we first compile the two component programs into STMs; a composite STM is then built from these two STMs, implementing the sequence semantics.

\begin{figure}[t]
 \centering
 \includegraphics[width=0.35\textwidth]{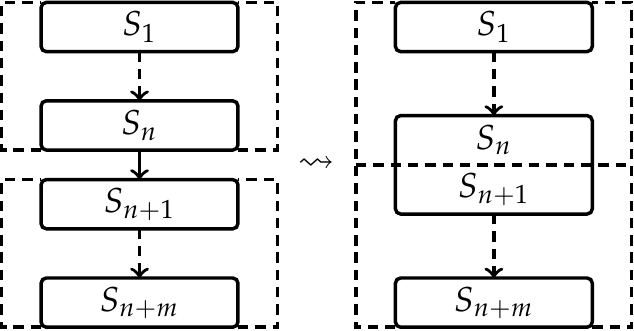}
 \caption{The compilation of sequence. A most straightforward way is shown on the left, and our compiler merges the states $S_n$ and $S_{n+1}$ and creates the STM on the right.}
 \label{fig:sequence}
\vspace{-2ex}\end{figure}

We illustrate the compilation in \autoref{fig:sequence}.
The left side is a straightforward way of compiling, and the right side is an optimized one produced by our compiler, with states $S_n$ and $S_{n+1}$ merged together.
This is because the separation of $S_n$ and $S_{n+1}$ is unnecessary:
every Palgol program describes an independent vertex-centric computation that does not rely on any incoming messages (according to our high-level model); correspondingly, our compilation ensures that the first superstep in the compiled program ignores the incoming messages.
We call this the \emph{message-independence} property.
Since $S_{n+1}$ is the beginning of the second Palgol program, it ignores the incoming messages, and therefore the barrier synchronization between $S_n$ and $S_{n+1}$ can be omitted.

\subsubsection{Compiling Iterations}

\begin{figure}[h]
 \centering
 \includegraphics[width=0.45\textwidth]{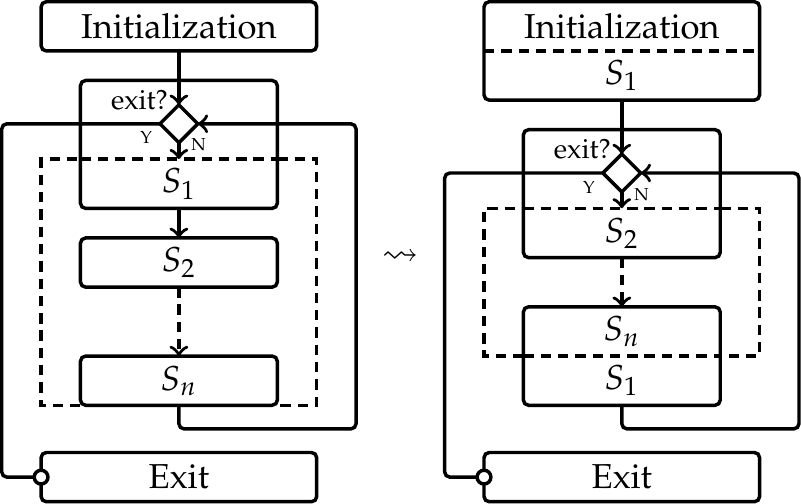}
 \caption{An STM for general iteration is shown on the left. The fusion optimization applies when the iteration body begins with a remote reading superstep ($S_1$), and yields the STM on the right.}
 \label{fig:fusion}
\vspace{-2ex}\end{figure}

Fixed-point iteration repeatedly runs a program enclosed by `\textbf{do}' and `\textbf{until} \ldots' until the specified fields stabilize.
To compile an iteration, we first compile its body into an STM, then we extend this STM to implement the fixed-point semantics.
The output STM is presented in \autoref{fig:fusion}, where the left one is generated by our general approach, and the right one performs the \emph{fusion optimization} when some condition is satisfied.

Let us start from the general approach on the left.
Temporarily ignoring the initialization state, the STM implements a while loop:
first, a check of the termination condition takes place right before the state $S_1$: if the termination condition holds, we immediately enters the state \textit{Exit}; otherwise we execute the body, after which we go back to the check.
The termination check is implemented by an OR aggregator to make sure that every vertex makes the same decision:
basically, every vertex determines whether its local fields are changed during a single iteration by storing the original values before~$S_1$, and sends the result (as a boolean) to the aggregator, which can then decide globally whether there exists any vertex that has not stabilized.
What remains is the initialization state, which guarantees that the termination check will succeed in the first run, turning the while loop into a do-until loop.

There is a chance to reduce the number of supersteps in the loop body of the iteration STM when the first state $S_1$ of the loop body is a remote reading superstep (see \autoref{sec:trans-step}).
In this case, as shown on the right side of \autoref{fig:fusion}, the termination check is moved to the beginning of the second state $S_2$, and then the state $S_1$ is duplicated and attached to the end of both the initialization state and $S_n$.
This transformation ensures that, no matter from where we reach the state $S_2$, we always execute the code in $S_1$ in the previous superstep to send the necessary messages.
With this property guaranteed, we can simply iterate $S_2$ to $S_n$ to implement the iteration, so that the number of supersteps inside the iteration is reduced.
The only difference with the left STM is that we execute an extra $S_1$ attached at the end of $S_n$ when we exit the iteration.
However, it still correctly implements the semantics of iteration:
the only action performed by a remote reading superstep is sending some messages;
although unnecessary messages are emitted, the Palgol program following the extra $S_1$ will ignore all incoming messages in its first state, as dictated by the message-independence property.

\subsection{Combiner Optimization}
\label{sec:combiner}

Combiners are a mechanism in Pregel that may reduce the number of messages transmitted during the computation.
Essentially, in a single superstep, if all the messages sent to a vertex are only meant to be consumed by a reduce-operator (e.g., sum or maximum) to produce a value on that vertex, and the values of the individual messages are not important, then the system can combine the messages intended for the vertex into a single one by that operator, reducing the number of messages that must be transmitted and buffered.

In Pregel, combiners are not enabled by default, since ``there is no mechanical way to find a useful combining function that is consistent with the semantics of the user's \emph{compute()} method''~\cite{pregel}.
However, Palgol's list comprehension syntax combines remote access and a reduce operator, and naturally represents such type of computation, which can potentially be optimized by a combiner.
A typical example is the SSSP program (line 7--8 in \autoref{fig:sssp-palgol}), where the distances received from the neighbors ($D[e.\mathbf{ref}]+e.\mathbf{val}$) are transmitted and reduced by the \textbf{minimum} operator.
Since the algorithm only cares about the minimum of the messages, and the compiler knows that nothing else is carried by the messages in that superstep, the compiler can automatically implement a combiner with the minimum operator to optimize the program.

\section{Experiments}
\label{sec:evaluation}

In this section, we evaluate the overall performance of Palgol and the state-merging optimisations introduced in the previous section.
We compile Palgol code to Pregel\plus\footnote{\url{http://www.cse.cuhk.edu.hk/pregelplus}}, which is an open-source implementation of Pregel written in C\plusplus.%
\footnote{Palgol does not target a specific Pregel-like system.
Instead, by properly implementing different backends of the compiler, Palgol can be transformed into any Pregel-like system, as long as the system supports the basic Pregel interfaces including message passing between arbitrary pairs of vertices and aggregators.}
We have implemented the following six graph algorithms on Pregel\plus's basic mode, which are:
\begin{itemize} 
 \item PageRank~\cite{pregel}
 \item Single-Source Shortest Path (SSSP)~\cite{pregel}
 \item Strongly Connected Components (SCC)~\cite{connectivity}
 \item Shiloach-Vishkin Connected Component Algorithm (S-V)~\cite{connectivity}
 \item List Ranking Algorithm (LR)~\cite{connectivity}
 \item Minimum Spanning Forest (MSF)~\cite{boruvka}
\end{itemize}
Among these algorithms, SCC, S-V, LR and MSF are non-trivial ones which contain multiple computing stages.
Their Pregel+ implementations are included in our repository for interested readers.

\begin{table}[t]
 \centering
 \caption{Datasets for Performance Evaluation}
 \label{tab:datasets}
 \begin{tabular}{c|c|c|c}
  \hline
  \textbf{Dataset} & \textbf{Type} & $|V|$ & $|E|$ \\
  \hline\hline
  Wikipedia & Directed & 18,268,992 & 172,183,984 \\
  \hline
  Facebook & Undirected & 59,216,214 & 185,044,032 \\
  \hline
  USA & Weighted & 23,947,347 & 58,333,344 \\
  \hline
  Random & Chain & 10,000,000 & 10,000,000 \\
  \hline
 \end{tabular}
\end{table}

We use 4 real-world graph datasets in our performance evaluation, which are listed in Table~\ref{tab:datasets}:
(1) Wikipedia\footnote{http://konect.uni-koblenz.de/networks/dbpedia-link}: the hyperlink network of Wikipedia;
(2) Facebook\footnote{https://archive.is/o/cdGrj/konect.uni-koblenz.de/networks/facebook-sg}: a friendship network of the Facebook social network;
(3) USA\footnote{http://www.dis.uniroma1.it/challenge9/download.shtml}: the USA road network;
(4) Random: a chain with randomly generated values.

The experiment is conducted on an Amazon EC2 cluster with 16 nodes (whose instance type is m4.large), each containing 2 vCPUs and 8G memory.
Each algorithm is run on the type of input graphs to which it is applicable (PageRank on directed graphs, for example) with 4 configurations, where the number of nodes changes from 4 to 16.
We measure the execution time for each experiment, and all the results are averaged over three repeated experiments.
The runtime results of our experiments are summarized in \autoref{tab:exec}.

\begin{table*}[t]
 \centering
 \caption{Comparison of Execution Time between Palgol and Pregel\protect\plus\ Implementation}
 \label{tab:exec}
 \resizebox{\textwidth}{!}{
 \begin{tabular}{c|c||c|c||c|c||c|c||c|c||c}
  \hline
  \multirow{2}{*}{\textbf{Dataset}} & \multirow{2}{*}{\textbf{Algorithm}} & \multicolumn{2}{c||}{4 nodes} & \multicolumn{2}{c||}{8 nodes} & \multicolumn{2}{c||}{12 nodes} & \multicolumn{2}{c||}{16 nodes} & \multirow{2}{*}{Comparison} \\
  \cline{3-10}
  & & Pregel\plus & Palgol & Pregel\plus & Palgol & Pregel\plus & Palgol & Pregel\plus & Palgol & \\
  \hline\hline
  \multirow{3}{*}{Wikipedia} & SSSP & 8.33 & 10.80 & 4.47 & 5.61 & 3.18 & 3.83 & 2.41 & 2.85 & 18.06\% -- 29.55\% \\
  \cline{2-11}
  & PageRank & 153.40 & 152.36 & 83.94 & 82.58 & 61.82 & 61.24 & 48.36 & 47.66 & -1.62\% -- 2.26\% \\
  \cline{2-11}
  & SCC & 177.51 & 178.87 & 85.87 & 86.52 & 61.75 & 61.89 & 46.64 & 46.33 & -0.66\% -- 0.77\% \\
  \hline
  Facebook & S-V & 143.09 & 142.16 & 87.98 & 86.22 & 67.62 & 65.90 & 58.29 & 57.49 & -2.53\% -- -0.65\% \\
  \hline
  Random & LR & 56.18 & 64.69 & 29.58 & 33.17 & 19.76 & 23.48 & 14.64 & 18.16 & 12.14\% -- 24.00\% \\
  \hline
  USA & MSF & 78.80 & 82.57 & 43.21 & 45.98 & 29.47 & 31.07 & 22.84 & 24.29 & 4.79\% -- 6.42\% \\
  \hline
 \end{tabular}}
\end{table*}

Remarkably, for most of these algorithms (PageRank, SCC, S-V and MSF), we observed highly close execution time on the compiler-generated programs and the manually implemented programs,
with the performance of the Palgol programs varying between a $2.53\%$ speedup to a $6.42\%$ slowdown.
The generated programs for PageRank and S-V are almost identical to the hand-written versions,
while some subtle differences exist in SCC and MSF programs.
For SCC, the whole algorithm is a global iteration with several iterative sub-steps, and the human written code can exit the outermost iteration earlier by adding an extra assertion in the code (like a \textbf{break} inside a \textbf{do ... until} loop).
Such optimization is not supported by Palgol currently.
For MSF, the human written code optimizes the evaluation of a special expression $D[D[u]] \shorteqq u$ to only one round of communication, while Palgol's strategy always evaluates the chain access $D[D[u]]$ using a request followed by and a reply step, and then compares the result with $u$.
These differences are however not critical to the performance.

For SSSP, we observed a slowdown up to $29.55\%$.
The main reason is that the human-written code utilizes Pregel's \textit{vote\_to\_halt()} API to deactivate converged vertices during computation;
this accelerates the execution since the Pregel system skips invoking the \emph{compute()} function for those inactive vertices, while in Palgol, we check the states of the vertices to decide whether to perform computation.
Similarly, we observed a $24\%$ slowdown for LR, since the human-written code deactivates all vertices after each superstep, and it turns out to work correctly.
While voting to halt may look important to efficiency, we would argue against supporting voting to halt as is, since it makes programs impossible to compose:
in general, an algorithm may contain multiple computation stages, and we need to control when to end a stage and enter the next; voting to halt, however, does not help with such stage transition, since it is designed to deactivate all vertices and end the whole computation right away.

\subsection{Effectiveness of Optimization}

\begin{table*}[t]
 \centering
 \caption{Comparison of the Compiler-Generated Programs Before/After Optimization}
 \label{tab:steps}
 \resizebox{\textwidth}{!}{
 \begin{tabular}{c|c||c|c|c||c|c|c}
  \hline
  \multirow{2}{*}{\textbf{Dataset}} & \multirow{2}{*}{\textbf{Algorithm}} & \multicolumn{3}{c||}{\# Supersteps} & \multicolumn{3}{c}{Execution Time} \\
  \cline{3-8}
  & & ~~Before~~ & ~~After~~ & Comparison & ~~Before~~ & ~~After~~ & Comparison \\
  \hline\hline
  \multirow{3}{*}{Wikipedia} & SSSP & 147 & 50 & -65.99\% & 5.36 & 2.85 & -46.83\% \\
  \cline{2-8}
  & PageRank & 93 & 32 & -65.59\% & 45.57 & 47.66 & 4.58\% \\
  \cline{2-8}
  & SCC & 3819 & 1278 & -66.54\% & 106.03 & 46.33 & -56.30\% \\
  \hline
  Facebook & S-V & 31 & 23 & -25.81\% & 52.37 & 57.49 & 9.78\% \\
  \hline
  Random & LR & 77 & 52 & -32.47\% & 17.54 & 18.16 & 3.51\% \\
  \hline
  USA & MSF & 318 & 192 & -39.62\% & 26.67 & 24.29 & -8.95\% \\
  \hline
 \end{tabular}}
\end{table*}

In this subsection, we evaluate the effectiveness of the ``state merging'' optimization mentioned in \autoref{sec:trans-iter}, 
by generating both the optimized and unoptimized versions of the code and executing them in the same configurations.
We use all the six graph applications in the previous experiment, and fix the number of nodes to 16.

The experiment results are shown in \autoref{tab:steps}.
From this table, we observed a significant reduction on the number of supersteps for all graph algorithms after optimization.
Then, SSSP and SCC are roughly twice faster than the unoptimized version, but for other algorithms, the optimization has relatively small effect on execution time.

First, the reduction of the number of supersteps in execution has a strong connection with the optimization results of the main iteration in these graph algorithms.
For applications containing only a simple iteration like PageRank and SSSP, we reduced nearly $2/3$ supersteps in execution, which is achieved by optimizing the three supersteps inside the iteration body into a single one.
Similarly, for SCC, S-V and LR, the improvement is around $2/3$, $1/4$ and $1/3$ due to the reduction of one or two superstep in the main iteration(s).
The MSF is a slightly complicated algorithm containing multiple stages, and we get an overall reduction of nearly $40\%$ supersteps in execution.

The effect of this optimization on execution time is however related to not only the number of supersteps that are reduced, but also the property of the applications.
On one hand, it reduces the number of global synchronizations so that the performance can be improved.
On the other hand, our optimization brings a small overhead, because we obtain a tighter loop body by unconditionally sending the necessary messages for the next iteration at the end of each iteration.
As a result, when exiting the loop, some redundant messages are emitted (although the correctness of the generated code is ensured).
In our experiments, SSSP and SCC are roughly twice faster after optimization, since they are not computational intensive, so that the number of global synchronization matters.

\section{Related Work}
\label{sec:related}

Google's Pregel \cite{pregel} proposed the vertex-centric computing paradigm, which allows programmers to think naturally like a vertex when designing distributed graph algorithms.
There are a bunch of open-source alternatives to the official and proprietary Pregel system, such as Apache Hama~\cite{hama}, Apache Giraph~\cite{giraph}, Catch the Wind~\cite{catchw}, GPS~\cite{gps}, GraphLab~\cite{graphlab}, PowerGraph~\cite{powergraph} and Mizan~\cite{mizan}.
This paper does not target a specific Pregel-like system.
Some graph-centric (or block-centric) systems like Giraph\plus\cite{thinkgraph} and Blogel~\cite{blogel} extends Pregel's vertex-centric approach by making the partitioning mechanism open to programmers, but it is still unclear how to optimize general vertex-centric algorithms (especially those complicated ones containing non-trivial communication patterns) using such extension.

Domain-Specific Languages (DSLs) are a well-known mechanism for describing solutions in specialized domains.
To ease Pregel programming, many DSLs have been proposed, such as Palovca~\cite{palovca}, s6raph~\cite{s6raph}, Fregel~\cite{fregel} and Green-Marl~\cite{green14}.
We briefly introduce each of them below.

Palovca~\cite{palovca} exposes the Pregel APIs in Haskell using a monad, and a vertex-centric program is written in a low-level way like in typical Pregel systems.
Since this language is still low-level, programmers are faced with the same challenges in Pregel programming, mainly having to tackle all low-level details.

At the other extreme, the s6raph system~\cite{s6raph} is a special graph processing framework with a functional interface.
It models a particular type of iterative vertex-centric computation by six programmer-specified functions, and can only express graph algorithms that contain a single iterative computation (such as PageRank and Shortest Path), whereas many practical Pregel algorithms are far more complicated.

A more comparable and (in fact) closely related piece of work is Fregel~\cite{fregel}, which is a functional DSL for declarative programming on big graphs.
In Fregel, a vertex-centric computation is represented by a pure step function that takes a graph as input and produces a new vertex state;
such functions can then be composed using a set of predefined higher-order functions to implement a complete graph algorithm.
Palgol borrows this idea in the language design by letting programmers write atomic vertex-centric computations called Palgol steps, and put them together using two combinators, namely sequence and iteration.
Compared with Fregel, the main strength of Palgol is in its remote access capabilities:
\begin{itemize}
 \item a Palgol step consists of local computation and remote updating phases, whereas a Fregel step function can be thought of as only describing local computation, lacking the ability to modify other vertices' states;
 \item even when considering local computation only, Palgol has highly declarative \textit{field access expressions} to express remote reading of arbitrary vertices, whereas Fregel allows only neighboring access.
\end{itemize}
These two features are however essential for implementing the examples in \autoref{sec:core}, especially the S-V algorithm.
Moreover, Palgol shows that Fregel's combinator-based design can benefit from Green-Marl's fusion optimizations~(\autoref{sec:trans-iter}) and achieve efficiency comparable to hand-written code.

Another comparable DSL is Green-Marl~\cite{green12}, which lets programmers describe graph algorithms in a higher-level imperative language.
This language is initially proposed for graph processing on the shared-memory model, and a ``Pregel-canonical'' subset of its programs can be compiled to Pregel.
Since it does not have a Pregel-specific language design, programmers may easily get compilation errors if they are not familiar with the implementation of the compiler.
In contrast, Palgol (and Fregel) programs are by construction vertex-centric and distinguish the current and previous states for the vertices, and thus have a closer correspondence with the Pregel model.
For remote reads, Green-Marl only supports neighboring access, so it suffers the same problem as Fregel where programmers cannot fetch data from an arbitrary vertex.
While it supports graph traversal skeletons like BFS and DFS, these traversals can be encoded as neighborhood access with modest effort, so it actually has the same expressiveness as Fregel in terms of remote reading.
Green-Marl supports remote writing, but according to our experience, it is quite restricted, and at least cannot be used inside a loop iterating over a neighbor list, and thus is less expressive than Palgol.

\section{Concluding Remarks}
\label{sec:conclusions}

This paper has introduced Palgol, a high-level domain-specific language for Pregel systems with flexible remote data access, which makes it possible for programmers to express Pregel algorithms that communicate over dynamic internal data structures.
We have demonstrated the power of Palgol's remote access by giving two representative examples, the S-V algorithm and the list ranking algorithm, and presented the key algorithm for compiling remote access.
Moreover, we have shown that Fregel’s more structured approach to vertex-centric computing can achieve high efficiency --- the experiment results show that graph algorithms written in Palgol can be compiled to efficient Pregel programs comparable to human written ones.

We expect Palgol's remote access capabilities to help with developing more sophisticated vertex-centric algorithms where each vertex decides its action by looking at not only its immediate neighborhood but also an extended and dynamic neighborhood.
The S-V and list ranking algorithms are just a start --- for a differently flavored example, graph pattern matching~\cite{graphpm} might be greatly simplified when the pattern has a constant size and can be translated declaratively as a remote access expression deciding whether a vertex and some other ``nearby'' vertices exhibit the pattern.
Algorithm design and language design are interdependent, with algorithmic ideas prompting more language features and higher-level languages making it easier to formulate and reason about more sophisticated algorithms.
We believe that Palgol is a much-needed advance in language design that can bring vertex-centric algorithm design forward.

\balance

%

\bibliographystyle{abbrv}
\bibliography{ref} 

\end{document}